\documentclass[10pt]{article}
\usepackage{amssymb,amsmath,amsthm,mathrsfs}
\newcommand{\R}{\mathbb{R}}
\newcommand{\tr}{\text{tr}}
\newcommand{\pd}{\partial}
\newcommand{\N}{\mathbb{N}}
\newcommand{\Sp}{\mathbb{S}}

\newcommand{\eps}{\varepsilon}

\newtheorem{Lemma}{Lemma}
\newtheorem{Theorem}{Theorem}

\newtheorem{Corollary}{Corollary}

\theoremstyle{definition}
\newtheorem{Remark}{Remark}
\newtheorem{ass}{Assumption}

\begin{document}

\title{\bf Spectral estimates for two-dimensional Schr\"odinger operators
  with application to quantum layers}

\author{
Hynek Kova\v{r}{\'\i}k, Semjon Vugalter and Timo Weidl
}
\date{
\begin{center} {\small 
Institute of Analysis, Dynamics
and Modeling, Universit\"at Stuttgart, 
PF 80 11 40, D-70569  Stuttgart, Germany.
}
\end{center}
}
%
%-------%
% TITLE %
%-------%
%------------------------------------------%
\maketitle

\begin{abstract}
\noindent
A logarithmic type Lieb-Thirring inequality for two-dimensional
Schr\"odinger operators is established. The result is applied to
prove spectral estimates on trapped modes in quantum layers.
\end{abstract}

%--------------

\section{Introduction}

It is well known that the sum of the moments of negative eigenvalues
$-\lambda_j$ of a one-dimensional Schr\"odinger operator $-\frac{d^2}{dx^2} -V$ 
can be estimated by 
\begin{equation} \label{Timo}
\sum_j\, \lambda_j^{\gamma} \, \, \leq \, \, L_{\gamma,1}\,
\int_{\R}\, V_+(x)^{\gamma+\frac 12} \, dx, \quad \gamma \geq \frac 12\, ,
\end{equation}
where $L_{\gamma,1}$ is a constant independent of $V$, see \cite{LT}, \cite{We}.
For $\gamma=\frac 12$ this bound has the correct weak coupling 
behavior, see \cite{Si}, and it also shows the correct Weyl-type asymptotics 
in the semi-classical limit. Moreover, \eqref{Timo} fails to hold whenever 
$\gamma < \frac 12$. The case $\gamma=\frac 12$ therefore represents certain 
borderline inequality in dimension one.

The situation is much less satisfactory in dimension two. The corresponding
two-dimensional Lieb-Thirring bound
\begin{equation} \label{Lieb-Thirring}
\sum_j\, \lambda_j^{\gamma} = \tr
\left(-\Delta-V\right)_-^{\gamma} \, \leq \, L_{\gamma,2}\,
\int_{\R^2}\, V_+(x)^{\gamma+1}\, dx
\end{equation}
holds for all $\gamma> 0$, \cite{LT}. Dimensional analysis shows
that here the borderline should be $\gamma=0$. However,
(\ref{Lieb-Thirring}) fails for $\gamma=0$, because
$-\Delta -V$ has at least one negative eigenvalue
whenever $\int\, V \geq 0$\, , see \cite{Si}. In addition, it was
shown in \cite{Si} that if $V$ decays fast enough, the operator
$-\Delta-\alpha V$  
has for small $\alpha$ only one eigenvalue which goes to zero exponentially 
fast:
\begin{equation} \label{simon}
\lambda_1 \sim e^{-4\pi(\alpha\int V)^{-1}}\, , \quad \alpha\to 0\, .
\end{equation}
It follows from \eqref{simon} that the optimal behavior for $\alpha\to 0$
cannot be reached in the power-like scale \eqref{Lieb-Thirring}, 
no matter how small $\gamma$ is, since the l.h.s.~decays faster than any 
power of $\alpha$. This means that in order to obtain a Lieb-Thirring type 
inequality with the optimal behavior in the weak coupling limit, 
one should introduce a different scale on the 
l.h.s.~of \eqref{Lieb-Thirring}.

\vspace{0.15cm}

In the present paper we want to find a two-dimensional analog of the
one-dimensional 
borderline inequality, which corresponds to $\gamma=\frac 12$ in
\eqref{Timo}. In other words, we want to establish an inequality
with the r.h.s.~proportional to $V$ and with the correct order of
asymptotics in weak and strong coupling regime. Obviously, we have
to replace the power function on the l.h.s.~of (\ref{Lieb-Thirring})
by a new function $F(\lambda)$, which will approximate identity as
close as possible. On the other hand, since $-\Delta-V$ has always
at least one eigenvalue, it is necessary that $F(0)=0$. Moreover,
equation \eqref{simon} shows that $F$ should grow from zero faster
than any power of $\lambda$, namely as $|\ln \lambda|^{-1}$. This
leads us to define the family of functions $F_s:(0,\infty)\to(0,1]$ by
\begin{equation}
\forall\, s>0 \qquad F_s(t) := \left\{
\begin{array}{l@{\quad \mathrm{} \quad }l}
 |\ln ts^2|^{-1} & 0< t \leq e^{-1}s^{-2}\, , \\
 &  \\
1 & t > e^{-1}s^{-2} \, .
\end{array}
\right.
\end{equation}
Notice that each $F_s$ is non decreasing and continuous and that
$F_s(t)\to 1$ point-wise as $s\to\infty$. Hence our goal is to
establish an appropriate estimate on the regularized counting
function
%\begin{equation} \label{sum}
$
\, \sum_j\, F_s(\lambda_j) \,
$
%\end{equation}
for large values of the parameter $s$.

\vspace{0.15cm}

Our main results is formulated in the next section. It turns out,
that $\, \sum_j\, F_s(\lambda_j) \,$ can be estimated by a sum of
two integrals, one of which includes a local logarithmic weight, see
Theorem \ref{2dim}. The inequality \eqref{first-ineq} established in
Theorem \ref{2dim} has the correct behavior for weak as
well as for strong potentials, see Remark \ref{rem1}. 
We also show that the logarithmic weight in \eqref{first-ineq} cannot be
removed, see Remark \ref{rem2}. 
Moreover, in Corollary \ref{corol} we obtain individual
estimates on eigenvalues of Schr\"odinger operators with slowly
decaying potentials. The proof of the main result, including two
auxiliary Lemmata, is then given in section \ref{proof}. In the
closing section \ref{applications} we apply Theorem \ref{2dim} to
analyze discrete spectrum of a Schr\"odinger operator corresponding
to quantum layers. The result established in 
section \ref{applications} may be regarded as two-dimensional analog of
Lieb-Thirring inequalities on trapped modes in quantum waveguides
obtained in \cite{EW}.

\section{Main results}
\noindent For a given $V$ we define the Schr\"odinger operator
\begin{equation}
-\Delta - V \quad \text{in}\quad L^2(\R^2)
\end{equation}
as the Friedrich extension of the operator associated with the quadratic form
\begin{equation} \label{qform}
Q_V[u] = \int_{\R^2}\, \left (|\nabla u|^2 -V|u|^2\right)\, dx \quad
\text{on} \quad C_0^{\infty}(\R^2)\, ,
\end{equation}
provided $Q_V$ is bounded from below. Throughout the paper we will suppose that
$V$ satisfies

\begin{ass} \label{basic}
The function $V(x)$ is such that
$
\sigma_{ess} (-\Delta-V) = [0,\infty)\, .
$
\end{ass}

Following notation will be used in the text. Given a self-adjoint
operator $T$, the number of negative eigenvalues, counting their
multiplicity, of $T$ to the left
of a point $-\nu$ is denoted by $N(\nu,T)$.  The symbol $\R_+$ stands
for the set $(0,\infty)$.
Moreover, as in \cite{LN} we define the space
$L^1(\R_+,L^p(\Sp^1))$ in polar coordinates $(r,\theta)$ in $\R^2$,
as the space of functions $f$ such that
\begin{equation} \label{L1p}
\|f\|_{L^1(\R_+, L^p(\Sp^1))} : = \int_0^{\infty}
\left(\int_0^{2\pi}
  |f(r,\theta)|^p\, d\theta\right)^{1/p}\, r\, dr \, < \infty \, .
\end{equation}
Finally, given $s>0$ we denote $B(s):=\{x\in \R^2\, :\, |x| < s \}$.
We then have

\begin{Theorem} \label{2dim}
Let $V\geq 0$ and $V\in L^1_{loc}(\R^2,|\ln|x||\, dx)$. Assume that
$V\in
L^1(\R_+,L^p(\Sp^1))$ for some $p>1$.
Then the quadratic form \eqref{qform} is bounded from below and
closable. The negative eigenvalues $-\lambda_j$ of the
operator associated with its closure satisfy the inequality
\begin{equation} \label{first-ineq}
\sum_j\, F_s(\lambda_j) \, \leq \, c_1\, \|V\, \ln
(|x|/s)\|_{L^1(B(s))}\, + c_p\, \|V\|_{L^1(\R_+, L^p(\Sp^1))}\,
\end{equation}
for all $s\in\R_+$. The constants $c_1$ and $c_p$ are independent of
$s$ and $V$.

\vspace{0.1cm}

\noindent
In particular, if $V(x) = V(|x|)$, then there exists a constant $C$, such that
\begin{equation} \label{radial}
\sum_j\, F_s(\lambda_j)  \, \leq \,  C \left(
\|V\ln(|x|/s)\|_{L^1(B(s))}\, + \|V\|_{L^1(\R^2)}\, \right)
\end{equation}
holds true for all $s\in\R_+$.
\end{Theorem}

\vspace{0.2cm}

\begin{Remark} \label{rem1}
Notice that the r.h.s.~of (\ref{first-ineq}) has the right order of
asymptotics in both weak and strong coupling limits.
Indeed, replacing $V$ by $\alpha V$ and assuming that
$V\in L^1(\R^2,(|\ln|x||+1)\, dx)$ it can be seen from the
definition of $F_s$ that
$$
\sum_j\, F_s(\lambda_j) \, \sim\, \alpha \, ,\quad
\alpha\to 0 \, \, \vee \, \,  \alpha\to\infty\, .
$$
For $\alpha\to 0$ this follows from
\eqref{simon}. For $\alpha\to\infty$ is the behavior of $
\sum_j\, F_s$ governed by the Weyl asymptotics for the counting
function:
\begin{equation} \label{sandwich}
N(e^{-1}s^{-2}, -\Delta-\alpha V) \, \leq \, \sum_j\, F_s(\lambda_j)
\, \leq \, N(0,-\Delta-\alpha V)\, .
\end{equation}
The latter is linear in $\alpha$ when $\alpha\to\infty$ provided
$V\in L^1(\R^2,(|\ln|x||+1)\, dx)$, see also Remark \ref{number}.
\end{Remark}

\begin{Remark} \label{rem2}
We would like to emphasize that $\sum_j\, F_s(\lambda_j)$
cannot be estimated only in terms of $\|V\|_{L^1(\R^2)}$. In particular,
the logarithmic term
in (\ref{first-ineq}) and (\ref{radial}) cannot be removed. This is due to 
the fact that there exist potentials $V\in
L^1(\R^2)$ with a strong 
local singularity, such that the semi-classical asymptotics of
$N(\nu,-\Delta-V)$ is non-Weyl for any $\nu >0$, \cite{BL}. Namely if we define
\begin{eqnarray}
V_\sigma(x) & = & r^{-2}\, |\ln r|^{-2}\, |\ln |\ln r||^{-1/\sigma},\quad r<
e^{-2},\quad
\sigma >1  \nonumber \\
V_\sigma(x) & = & 0, \qquad r \geq e^{-2} \, ,
\end{eqnarray}
where $r=|x|$, then $V_\sigma\in L^1(\R^2)$ for all $\sigma>1$, but
\begin{equation} \label{counter}
N(\nu,\, -\Delta-\alpha V_\sigma)\, \sim \, \alpha^\sigma\, \qquad
\alpha\to\infty \, , \quad \forall\ \nu >0\, ,
\end{equation}
see \cite[Sec.~6.5]{BL}. If \eqref{radial} were true with
the logarithmic factor removed, it would be in obvious contradiction
with (\ref{sandwich}) and \eqref{counter}. Moreover, the asymptotics
\eqref{counter} 
remains valid also if the singularity of $V$ is not placed at zero,
but at some other point. This shows that the condition $p>1$ in Theorem
\ref{2dim} is necessary. 
\end{Remark}

\begin{Remark} \label{rem3}
The non-Weyl asymptotics of $N(0,-\Delta-\alpha V)$
can also occur for potentials which have no singularities, but which decay at 
infinity too slowly, so that the associated eigenvalues accumulate at zero.
For example, if
\begin{eqnarray} \label{accum}
V^{\Phi}_\sigma(x) & = & \Phi(\theta)\, r^{-2}\, (\ln r)^{-2}\, (\ln \ln
r)^{-1/\sigma},\quad r> e^{2},\quad \sigma >1  \nonumber \\
V^\Phi_\sigma(x) & = & 0, \qquad r \leq e^{2} \, ,
\end{eqnarray}
then 
$$
N(0,-\Delta-\alpha V^\Phi_\sigma)\sim \alpha^\sigma\, ,
$$ 
see \cite{BL}. In this case, however, Theorem \ref{2dim} says that
the eigenvalues accumulating at zero
are small enough so that their total contribution to $\sum_j\,
F_s(\lambda_j)$ grows at most linearly in $\alpha$. More exactly, inequality
\eqref{first-ineq} gives the following estimate:

\end{Remark}

\begin{Corollary} \label{corol}
Let $\Phi\in L^p(0,2\pi)$ for some $p>1$. Let $V$ satisfy the
assumptions of Theorem \ref{2dim} and suppose that
$$
V(x)-V^{\Phi}_\sigma(x) = o\left(V^{|\Phi|}_\sigma(x)\right),\quad
|x|\to\infty \, ,
$$
where $V^{\Phi}_\sigma(x)$ is defined by \eqref{accum}. Denote
$n(\alpha)= N(0,-\Delta-\alpha V)$ and let $-\lambda_{n(\alpha)}$ be the
largest eigenvalue of $-\Delta-\alpha V$. Then, for any fixed $s>0$ 
there exists a constant $c_s>0$ such that for $\alpha$ large enough we
have 
\begin{equation} \label{estimate}
\lambda_{n(\alpha)} \leq \, s^{-2}\, \exp(-c_s\, \alpha^{\sigma-1})\, .
\end{equation}
\end{Corollary}

\begin{proof}
Inequality \eqref{first-ineq} shows that $\sum_j\, F_s(\lambda_j) \leq
c'_s \alpha$ for some $c'_s$. In particular, this implies
\begin{equation} \label{individual}
j\, F_s (\lambda_j) \, \leq \, c'_s\, \alpha\, ,\quad \forall\, j\, .
\end{equation}
On the other hand, from \cite[Prop.~6.1]{BL} follows that $n(\alpha) \geq
\tilde{c}\, \alpha^\sigma$ for some $\tilde{c}$ and $\alpha$ large enough.
An application of the inequality 
\eqref{individual} with $j=n(\alpha)$ then yields
\eqref{estimate}.
Analogous estimates for $\lambda_{n(\alpha)-k}\, ,k\in\N$ can be
obtained by an obvious modification.
\end{proof}

\section{Proof of Theorem \ref{2dim}}
\label{proof} We prove the inequality \eqref{first-ineq} for continuous
potentials with compact support.
The general case then follows by approximating $V$
by a sequence of continuous compactly supported functions and using a
standard limiting argument in \eqref{first-ineq}.

\vspace{0.15cm}
As usual in the borderline situations, the method of \cite{LT} cannot be 
directly applied and a different strategy is needed.
We shall treat the operator $-\Delta-V$ separately on the space of
spherically symmetric functions in $L^2(\R^2)$  and on its
orthogonal complement. To this end we define the corresponding
projection operators:
$$
(Pu)(r) = \frac{1}{2\pi}\, \int_0^{2\pi} u(r,\theta)\, d\theta\,
,\quad Qu = u-Pu\, ,\quad u\in L^2(\R^2)\, .
$$
Since $P$ and $Q$ commute with $-\Delta$,
the variational principle says that for each $a>1$ the operator inequality
\begin{equation} \label{variation}
-\Delta - V \geq P\, (-\Delta-(1+a^{-1})\, V)\, P +Q\,
(-\Delta-(1+a)\, V)\, Q
\end{equation}
holds. Let us denote by $-\lambda_j^P$ and $-\lambda_j^Q$ the
non decreasing sequences of
negative eigenvalues of the operators $P\ (-\Delta-(1+a^{-1})\, V)\, P$ and
$Q\, (-\Delta-(1+a\, V)\, Q$ respectively. Clearly we have
\begin{equation} \label{upperestimate}
\sum_j\, F_s(\lambda_j) \leq \sum_j\, F_s(\lambda^P_j) +\sum_j\,
F_s(\lambda^Q_j) \, .
\end{equation}
We are going to find appropriate bounds on the two terms on
the r.h.s.~of \eqref{upperestimate} separately. First we note that
$P\ (-\Delta-(1+a^{-1})\, V)\, P$ is unitarily equivalent to the operator
\begin{equation} \label{halfline}
h = -\frac{d^2}{d r^2}\, -\frac{1}{4r^2}\, - W(r) = h_0-W(r) \quad
\text{in} \quad L^2(\R_+)
\end{equation}
with the Dirichlet boundary condition at zero and with the potential
\begin{equation} \label{average}
W(r) = \frac{1+a}{2\pi a}\, \int_0^{2\pi}\, V(r,\theta)\,
d\theta\, .
\end{equation}
More precisely, $h$ is associated with the closure of the quadratic form
\begin{equation} \label{qform-hlafline}
q[\varphi] = \int_{\R_+}\, \left (|\varphi'|^2
-W|\varphi|^2\right)\, r\, dr \quad \text{on} \quad C_0^\infty(\R_+)\, .
\end{equation}
We start with the estimate on the lowest eigenvalue of $h$.

\begin{Lemma} \label{groundstate}
Let $V$ be continuous and compactly supported and let $W$ be given by
\eqref{average}. Denote by
$-\lambda^P_1$ the lowest eigenvalue of the operator $h$. Then
there exists a constant $c_2$, independent of $s$, such that
\begin{equation}
F_s(\lambda^P_1) \, \leq \, c_2\, \int_0^{\infty}\, W(r)\, r
\left(1+\chi_{(0, s)}(r) \, |\ln r/s|\, \right) \, dr\, .
\end{equation}
holds true for all $s\in\R_+$.
\end{Lemma}

\begin{proof}
From the Sturm-Liouville theory we find the Green function of the
operator $h_0$ at the point $-\kappa^2$:
$$
G_0(r,r',\kappa) := \left\{
\begin{array}{l@{\quad \mathrm{} \quad }l}
 \sqrt{rr'}\, I_0(\kappa r)\, K_0(\kappa r')
 & 0\leq r \leq r' < \infty,  \\
 &  \\
\sqrt{rr'}\, I_0(\kappa r')\, K_0(\kappa r) & 0\leq r' < r < \infty
\, ,
\end{array}
\right.
$$
where $I_0, K_0$ are the modified Bessel functions, see \cite{AS}.
The Birman-Schwinger principle tells us that if for certain value of
$\kappa$ the trace of the operator
$$
K(\kappa): = \sqrt{W}\,(h_0+\kappa^2)^{-1}\, \sqrt{W}
$$
is less than or equal to $1$, then the inequality $\lambda^P_1 \leq
\kappa^2$ holds. Taking into account the continuity of $W$, this implies
\begin{equation}  \label{trace}
\int_0^{\infty}\, r\, I_0\left(\sqrt{\lambda^P_1}\, r\right )\,
K_0\left(\sqrt{\lambda^P_1}\, r\right)\, W(r)\, dr \geq 1\, .
\end{equation}
Now we introduce the
substitutions $\tau =s\, \sqrt{\lambda^P_1}$,\, $t= s^{-1} r$ and recall that
$I_0(0)=1$ while $K_0$ has a logarithmic singularity at zero, see
\cite[Chap.9]{AS}. We thus find out that
$$
F_1\left(\tau^2 \right)\, I_0(\tau t)\, K_0(\tau t) \, \leq \, c_2\,
\left(1+ \chi_{(0,1)}(t)\, |\ln t| \right)\, ,\quad \forall
\tau\geq 0\, ,
$$
where $c_2$ is a suitable constant independent of $\tau$. Here we have
used the fact that
\begin{equation} \label{bessel}
\left |I_0(z)\, K_0(z)\right | \leq \text{const} \quad \forall\, z\geq
1\, ,
\end{equation}
see \cite{AS}. Finally, we multiply both sides of inequality (\ref{trace}) by
$F_s(\lambda^P_1)$ and note that
$$
F_s(\lambda^P_1)= F_s\left(\tau^2/s^2 \right)=F_1\left(\tau^2 \right)\, .
$$
The proof is complete.
\end{proof}

\noindent Next we estimate the higher eigenvalues of $h$.

\begin{Lemma} \label{dirichlet}
Under the assumptions of Lemma \ref{groundstate} there exists a constant
$c_3$ such that
$$
\sum_{j\geq 2}\, F_s(\lambda^P_j)\,  \leq\,
\int_0^{s}\, W(r)\, r\, \left |\ln r/s \right |\, dr
+ c_3\, \int_s^{\infty}\, W(r)\, r\, dr ,  \quad \forall\, s\in\R_+ \, .
$$
\end{Lemma}

\begin{proof}
Let us introduce the auxiliary operator
\begin{equation} \label{aux}
h_d = -\frac{d^2}{d r^2}\, -\frac{1}{4r^2}\, -W(r) \quad
\text{in}\quad  L^2(\R_+)
\end{equation}
subject to the Dirichlet boundary conditions at zero and at
the point $s$.
Let $-\mu_j$ be the non decreasing sequence of negative eigenvalues of
$h_d$. Since imposing
the Dirichlet boundary condition at $s$ is a rank one
perturbation, it follows from the variational principle that
\begin{equation} \label{upperbound}
\sum_{j\geq 2} \, F_s(\lambda^P_j)\, \leq\, \sum_{j\geq 1}\,
F_s(\mu_j)\, .
\end{equation}
Moreover, $h_d$ is unitarily equivalent to the orthogonal sum
$h_1\oplus h_2$, where
\begin{align} \label{decomp}
h_1 & = h_{1,0} -W(r) = -\frac{d^2}{d r^2}\, -\frac{1}{4r^2}\, -W(r) \quad
\text{in}\quad L^2(0,s) \nonumber \\
h_2 & = h_{2,0} -W(r) = -\frac{d^2}{d r^2}\, -\frac{1}{4r^2}\, -W(r) \quad
\text{in}\quad L^2(s,\infty) \nonumber
\end{align}
with Dirichlet boundary conditions at $0$ and $s$. Keeping in mind that
$F_s\leq 1$ we will estimate \eqref{upperbound} as follows:
\begin{equation} \label{h2-bound}
\sum_{j}\, F_s(\mu_j)\,  \leq\,  N(0,h_1)+\sum_{j}\, F_s(\mu'_j)\, ,
\end{equation}
where $-\mu'_j$ are the negative eigenvalues of $h_2$. To continue
we calculate the diagonal elements of the Green functions of the free
operators $h_{1,0}$ and $h_{2,0}$.
Similarly as in the proof of Lemma \ref{groundstate} we get
\begin{eqnarray} \label{green}
G_1(r,r,\kappa) & = &
 r\, I_0(\kappa r)\left(K_0(\kappa
   r)+\beta_s^{-1}(\kappa)I_0(\kappa
   r)\right)\qquad  0\leq r \leq s  \nonumber \\
G_2(r,r,\kappa) & = &
 r\, K_0(\kappa r)\left(I_0(\kappa r)+\beta_s(\kappa)K_0(\kappa
   r)\right) \qquad
  \, \, s\leq r <\infty\, ,
\end{eqnarray}
where
$$
\beta_s(\kappa) = -\frac{I_0(\kappa s)}{K_0(\kappa s)}\, .
$$
The Birman-Schwinger principle thus gives us the following
estimates on the number of eigenvalues of $h_1$ and $h_2$ to the left
of the point $-\kappa^2$:
\begin{equation} \label{bargman}
N(\kappa^2,h_1) \leq  \int_0^{s}\, G_1(r,r,\kappa)\, W(r)\, dr,
\quad N(\kappa^2,h_2) \leq  \int_{s}^{\infty}\, G_2(r,r,\kappa)\,
W(r)\, dr\, .
\end{equation}
Passing to the limit $\kappa\to 0$ and using the asymptotic
behavior of the Bessel functions $I_0$ and $K_0$, \cite{AS}, we
find out that for any fixed $r$ holds the identity
\begin{equation} \label{limit}
 \lim_{\kappa\to 0} G_1(r,r,\kappa) = \lim_{\kappa\to
0} G_2(r,r,\kappa) = r\, \left |\ln r/s \right |\,
\end{equation}
The assumption on $W$ and the dominated
convergence theorem then allow us to interchange the limit $\kappa\to 0$
with the integration in \eqref{bargman} to obtain
\begin{equation} \label{number-h_d}
N(0,h_1) \leq \int_0^{s}\, r\, \left |\ln r/s \right |\, W(r)\, dr
\, .
\end{equation}
This estimates the first term in \eqref{h2-bound}.
In order to find an upper bound on the second term in
\eqref{h2-bound}, we employ the formula
\begin{equation} \label{LT-formula}
\sum_{j}\, F_s(\mu'_j)\, = \int_0^{\infty}\, F'_s(t)\, N(t,h_2)\,
dt\, ,
\end{equation}
see \cite{LT}. Using \eqref{bargman}, the substitution $t\to
t^2\, $ and the Fubini theorem we get
$$
\sum_{j}\, F_s(\mu'_j)\, \leq \frac 12\, \int_s^{\infty} W(r)
\int_0^{e^{-1/2}s^{-1}}\, \frac{G_2(r,r,t)}{t\, (\ln ts)^2}\, \,
dt\, dr\, .
$$
In view of \eqref{green} it suffices to show that the integral
\begin{equation} \label{integral}
\int_0^{e^{-1/2}s^{-1}}\, \frac{K_0(tr)\left(I_0(tr)+
\beta_s(t)K_0(tr)\right)}{t\, (\ln ts)^2}\, \,
dt
\end{equation}
is uniformly bounded for all $s>0$ and $r\geq s$. The
substitutions $r=sy,\, t= \tau/s$ transform \eqref{integral} into
\begin{equation} \label{int-g}
g(y):= \int_0^{e^{-1/2}}\, \frac{K_0(\tau y)\left(I_0(\tau
    y)+\beta_1(\tau)K_0(\tau y)\right)}{\tau\, (\ln \tau)^2}\, \,
d\tau\, ,\quad y\in [1,\infty)\, .
\end{equation}
Since $g$ is continuous, due to the continuity of Bessel functions,
and $g(1)=0$, it is enough 
to check that $g(y)$ remains bounded as $y\to\infty$. Moreover, the
inequality
$$
(u, (h_{2,0}+t_1)^{-1}\, u) \leq (u,
(h_{2,0}+t_2)^{-1}\, u) \quad \forall\, 
\, 0\leq t_2\leq t_1\, \, ,\forall\, u\in L^2(s,\infty)
$$
shows that $G_2(r,r,t)$, the diagonal element of the integral kernel
of $(h_{2,0}+t^2)^{-1}$, is non increasing in $t$ for each $r\geq s$.
Equations \eqref{green} and \eqref{limit} then imply 
\begin{align}
\int_0^{y^{-1}}\, \frac{K_0(\tau y)\left(I_0(\tau
    y)+\beta_1(\tau)K_0(\tau y)\right)}{\tau\, (\ln \tau)^2}\, \,
d\tau\,  \leq \,
\ln y\, \int_0^{y^{-1}}\, \frac{d\tau}{\tau\,
  (\ln \tau)^2}\, =1  \nonumber\, .
\end{align}
On the other hand, when $\tau \in [y^{-1},e^{-1/2}]$, 
it can be seen from \eqref{bessel} and from the behavior of $I_0,K_0$ in
the vicinity of zero, see \cite{AS}, that
$$
\left |K_0(\tau y)\left(I_0(\tau y)+\beta_1(\tau)K_0(\tau
    y)\right)\right | \leq \text{const}
$$
uniformly in $y$. Equation \eqref{LT-formula} thus yields
$$
\sum_{j}\, F_s(\mu'_j)\, \leq c_3\, \int_s^{\infty}\, W(r)\, r\, dr
\quad \forall\, s\in\R_+\, ,
$$
where $c_3$ is independent of $s$.
Together with \eqref{upperbound}, \eqref{h2-bound}
and \eqref{number-h_d} this completes the proof.
\end{proof}
\noindent From equation \eqref{average}, Lemma \ref{groundstate} and Lemma
\ref{dirichlet} we conclude that
$$
\sum_j\, F_s(\lambda^P_j) \,  \leq  (c_2+1)\, \|V\ln
(|x|/s)\|_{L^1(B(s))}\, + c_3\, \|V\|_{L^1(\R^2)}\, .
$$

\vspace{0.15cm}

\noindent Let us now turn to the second term on the r.h.s.~of
\eqref{upperestimate}. The key  ingredient in estimating this
contribution will be the result of Laptev and Netrusov obtained in \cite{LN}.
We make use of the estimate
$$
\sum_j\, F_s(\lambda^Q_j) \,  \leq N\left(0,Q(-\Delta-(1+a)\,
V)Q\right)
$$
and of the Hardy-type inequality
\begin{equation} \label{hardy}
Q\, (-\Delta)\, Q \, \geq \, Q\, \frac{1}{|x|^2}\, Q\, ,
\end{equation}
which holds in the sense of quadratic forms on
$C_0^\infty(\R^2)$, see \cite{BL}.
For any $\eps\in (0,1)$ we thus get the lower bound
\begin{equation}
Q\, (-\Delta-(1+a)\, V)\, Q \geq
(1-\eps)\, Q\left(-\Delta+\frac{\eps}{1-\eps}\,
  \frac{1}{|x|^2}\, -\frac{1+a}{1-\eps}\, V\right)\, Q\, ,
\end{equation}
which implies
\begin{equation}
N\left(0,Q\, (-\Delta-(1+a)\, V)\, Q\right) \leq
N\left(0,-\Delta+\frac{\eps}{1-\eps}\,
  \frac{1}{|x|^2}\, -\frac{1+a}{1-\eps}\, V\right) \, .
\end{equation}
The last quantity can be estimated using \cite[Thm.1.2]{LN},
which says that
\begin{equation} \label{laptev-netrusov}
N\left(0,-\Delta+\frac{\eps}{1-\eps}\,
  \frac{1}{|x|^2}\, -\frac{1+a}{1-\eps}\, V\right)  \, \leq \,
\tilde{c}_p\, \|V\|_{L^1(\R_+, L^p(\Sp^1))}\, .
\end{equation}
for some constant $\tilde{c}_p$ that also depends on $\eps$ and $a$.
In order to conclude the proof of (\ref{first-ineq}) we note that by
the H\"older inequality
$$
\|V\|_{L^1(\R^2)} \, \leq \text{const}\, \|V\|_{L^1(\R_+,
  L^p(\Sp^1))}\, .
$$

To show that the quadratic form \eqref{qform} is semi-bounded from
below we note that inequality (\ref{first-ineq}) says that
there are only finitely many eigenvalues of $-\Delta-V$ below
$-e^{-1}\, s^{-2}$. Let $-\Lambda_V$ be the minimum of those. Then
$$
Q_V[u]\, \geq \, -\Lambda_V\, \|u\|_{L^2(\R^2)} \quad \forall\, u\in
C_0^{\infty}(\R^2)\, .
$$
The proof of Theorem \ref{2dim} is now complete.

\begin{Remark} \label{number}
As a corollary of the proof of Theorem \ref{2dim} we immediately obtain
\begin{equation}
N(0,-\Delta-V) \, \leq \, 1 + \text{const}
\left(\|V\ln|x|\|_{L^1(\R^2)}+
\|V\|_{L^1(\R_+, L^p(\Sp^1))} \right)\, ,
\end{equation}
which agrees with \cite[Thm.3]{S}.
\end{Remark}
\begin{Remark} 
Lieb-Thirring inequalities for the operator $h=h_0-W$ in the form
$$
\tr\, (h_0-W)_-^\gamma\, \leq  \, C_{\gamma,a}\, \int_{\R_+}\,
W(r)_+^{\gamma+\frac{1+a}{2}}\, r^a\, dr, \quad \gamma >0\, , \, a\geq 1
$$
have been recently established in \cite{EF}. 
\end{Remark}

\section{Application}
\label{applications}
In this section we consider a model of quantum layers.
It concerns a conducting plate $\Omega=\R^2\times(0,d)$
with an electric potential $V$. We will consider the shifted 
Hamiltonian
\begin{equation} \label{hamiltonian}
H_V = -\Delta_\Omega -V -\frac{\pi^2}{d^2}\, \quad \text{in} \quad
L^2(\Omega)\, , 
\end{equation}
with Dirichlet boundary conditions at $\pd\Omega$, which is associated with
the closed quadratic form
\begin{equation} \label{form-V}
\int_\Omega \left(|\nabla u|^2 -V |u|^2-\frac{\pi^2}{d^2}\,
  |u|^2\right)\, dx \quad \text{on}\quad H_0^1(\Omega)\, .
\end{equation}
We assume that for each $x_3\in (0,d)$ the function $V(\cdot\, ,\cdot
,x_3)$ satisfies Assumption \ref{basic}. Without loss of generality we assume
that $V\geq 0$, otherwise we replace $V$ by its positive part.

\noindent The essential spectrum of the Operator $H_V$  
covers the half line $\left[0
,\infty\right)$.
Let us denote by $-\tilde{\lambda}_j$ the
non decreasing sequences of negative eigenvalues of $H_V$.
For the sake of brevity we choose $s=1$ and prove

\begin{Theorem} \label{main1} Assume that $V\in L^{3/2}(\Omega)$ and
that
$$
\tilde{V}(x_1,x_2) = \frac 2d\, \int_0^d V(x_1,x_2,x_3)\,
\sin^2\left(\frac{\pi\, x_3}{d}\right)\, dx_3
$$
satisfies the assumptions of Theorem \ref{2dim} for some $p>1$. Then
there exist positive constants $C_1,C_2, C_3(p)$ such that
\begin{align} \label{mainineq1}
\sum_j F_1(\tilde{\lambda}_j)  \,  \leq\, &\, \,  C_1\,
\|\tilde{V} \ln (x_1^2+x_2^2)\|_{L^1(B(1))} \,
+ C_3(p)\,
\|\tilde{V}\|_{L^1(\R_+, L^p(\Sp^1))}\,  \nonumber \\
& + C_2\|V^{3/2}\|_{L^1(\Omega)}\, .
\end{align}
\end{Theorem}

\begin{Remark}
Notice that (\ref{mainineq1}) has the right asymptotic behavior in
both weak and strong coupling limits. Namely, in the weak coupling
limit the r.h.s. is dominated by the term linear in $V$, while in
the strong coupling limit prevails the term proportional to
$V^{3/2}$. In this sense our result is similar to the Lieb-Thirring 
inequalities on trapped modes in quantum wires obtained in \cite{EW}.
\end{Remark}

\begin{proof}[Proof of Theorem \ref{main1}]

Let $\nu_k= k^2 \pi^2/d^2,\, k\in\N$ be the eigenvalues of the Dirichlet
Laplacian on $(0,d)$ associated with the normalized eigenfunctions
$$
\phi_k(x_3) = \sqrt{\frac{2}{d}}\, \,
\sin\left(\frac{k\, \pi x_3}{d}\right)\, .
$$
Moreover, define
$$
R = (\phi_1,\cdot )\, \phi_1,\quad S = \Bbb I -R\, .
$$
By the same variational argument used in the previous section we
obtain the inequality
\begin{equation}
H_V \geq R\, (-\Delta_\Omega-\nu_1-2V)\, R + S\,
(-\Delta_\Omega-\nu_1-2V)\, S\, . 
\end{equation}
The latter implies
\begin{equation} \label{auxiliary}
\sum_j F_1(\tilde{\lambda}_j)\, \leq \sum_j\, F_1(\tilde{\mu}_j)
+N(0,\, S\, (-\Delta_\Omega-\nu_1-2V)\, S)\, ,
\end{equation}
where $-\tilde{\mu}_j$ are the negative eigenvalues of
$R\, (-\Delta_\Omega-\nu_1-2V)\, R$.
Since
$$
R\, (-\Delta_\Omega-\nu_1-2V)\, R = (-\pd_{x_1}^2 -\pd_{x_2}^2
-2\tilde{V})\otimes\, R\, ,
$$
the first term on the r.h.s. of (\ref{auxiliary}) can be
estimated using (\ref{first-ineq}) as follows:
\begin{equation}
\sum_j\, F_1(\tilde{\mu}_j) \, \leq \,
C_1\, \|\tilde{V}_1 \ln (x_1^2+x_2^2)\|_{L^1(\R^2)} \,
+ C_3(p)\, \|\tilde{V}\|_{L^1(\R_+, L^p(\Sp^1))}\, .
\end{equation}
As for the second term, we note that
\begin{align}
S\, (-\pd_{x_3}^2-\nu_1)\, S & = \sum_{k=2}^{\infty} (\nu_k-\nu_1)\,
(\phi_k,\cdot)\, \phi_k \geq \sum_{k=2}^{\infty}
\frac{\nu_2-\nu_1}{\nu_2}\,
\nu_k\, (\phi_k,\cdot)\, \phi_k \nonumber \\
& = \frac 34\, S\, (-\pd_{x_3}^2)\, S \nonumber
\end{align}
holds true in the sense of quadratic forms on $C_0^{\infty}(0,d)$, which
implies the estimate
$$
S\, (-\Delta_\Omega-\nu_1-2V)\, S\,  \geq \, \frac 34\, S\,
\left(-\Delta_\Omega-\frac 83\, 
  V\right)\, S\, .
$$
Using the variational principle and the Cwickel-Lieb-Rosenblum
inequality, \cite{C,L,R}, 
we thus arrive at
\begin{equation*}
N(0,S\, (-\Delta_\Omega-\nu_1-2V)\, S) \, \leq\,
 N\left(0,\, -\Delta_\Omega-\frac 83\, V\right)
\, \leq \, C_2 \int_{\Omega} V^{3/2}\, .
\end{equation*}
In view of \eqref{auxiliary} this concludes the proof.

\end{proof}

\section*{Acknowledegement}
We would like to thank Eliot Lieb for useful comments. The
support from the DFG grant WE 1964/2 is gratefully acknowledged.

%--------------%
% BIBLIOGRAPHY %
%--------------%
%
%\newpage
%\addcontentsline{toc}{section}{References}
%\bibliography{bib}
%\bibliographystyle{amsplain}
%
\providecommand{\bysame}{\leavevmode\hbox to3em{\hrulefill}\thinspace}
\providecommand{\MR}{\relax\ifhmode\unskip\space\fi MR }
\providecommand{\MRhref}[2]{%
  \href{http://www.ams.org/mathscinet-getitem?mr=#1}{#2}
}
\providecommand{\href}[2]{#2}
\end{document}